\RequirePackage{fix-cm}
\documentclass[smallcondensed]{svjour3}     
\smartqed  
\usepackage{graphicx}
\usepackage{amsmath}

\newcommand{\myskip}[1]{}

\newcommand{\tr}{{\rm tr}\,}

\newcommand{\minfty}{{-\infty}}

\newcommand{\vq}{{\bf q}}

\newcommand{\gam}{\tau}

\newcommand{\ccdot}{\hspace{-0.5mm}\cdot\hspace{-0.5mm}}

\newcommand{\vB}{{\bf B}}
\newcommand{\vE}{{\bf E}}

\newcommand{\vA}{{\bf A}}
\newcommand{\vC}{{\bf C}}

\newcommand{\vJ}{{\bf J}}

\newcommand{\vL}{{\bf L}}

\newcommand{\vS}{{\bf S}}

\newcommand{\cE}{{\cal E}}

\newcommand{\vk}{{\bf k}}
\newcommand{\vn}{{\bf n}}
\newcommand{\vp}{{\bf p}}
\newcommand{\vr}{{\bf r}}

\newcommand{\p}{{\partial}}
\newcommand{\veen}{{\bf 1}}
\newcommand{\vnul}{{\bf 0}}
\newcommand{\A}{{\bf A}}

\renewcommand{\p}{\partial}

\newcommand{\BEQ}{\begin{eqnarray}}
\newcommand{\EEQ}{\end{eqnarray}}
\newcommand{\BEA}{\begin{eqnarray}}
\newcommand{\EEA}{\end{eqnarray}}

\newcommand{\nn}{\nonumber }
\renewcommand{\d}{{{\rm d}}}
\newcommand{\eps}{\varepsilon}
\newcommand{\om}{\omega}

\newcommand{\half}{\frac{1}{2}}

\begin{document}

\title{Simulation of the hydrogen ground state in Stochastic Electrodynamics-2}
\subtitle{Inclusion of relativistic corrections}


\titlerunning{H-atom in Stochastic Electrodynamics}

\author{Theodorus M. Nieuwenhuizen \and Matthew T. P.  Liska}


\authorrunning{ Nieuwenhuizen and Liska}

\institute{
T. M. Nieuwenhuizen \at 
 Institute for Theoretical Physics, 
P.O. Box 94485, 1098 XH Amsterdam, the Netherlands  
 \emph{Second affiliation:
          International Institute of Physics, UFRG,
 Av. O. Gomes de Lima, 1722, 
59078-400  Natal-RN, Brazil}
          \and M. T. P. Liska \at Institute for Theoretical Physics
           P.O. Box 94485, 1098 XH Amsterdam, the Netherlands }

\date{Received: date / Accepted: date}

\maketitle

\begin{abstract}
In a recent paper the authors studied numerically the hydrogen ground state in stochastic electrodynamics (SED) within the the non-relativistic approximation.  
In quantum theory the leading non-relativistic corrections to the ground state energy dominate the Lamb shift related to the photon cloud that should cause
the quantum-like behaviour of SED.  The present work takes these corrections into account in the numerical modelling. It is found that they have little effect;
the self-ionisation that occurs without them remains present.  It is speculated that the point-charge approximation for the electron is the cause of the failure.
\end{abstract}

\section{Introduction}
\label{intro}
Stochastic Electrodynamics (SED) has been put forward as a candidate for a hidden variables theory that underlies quantum mechanics,
see the books \cite{delaPenaCettobook,delaPenaCettoValdesbook}. Such theories are not ruled out by Bell's theorem, since that
has an irreparable fatality, the contextuality loophole \cite{TheoBellFoP}. 

The derivation of the Schr\"odinger equation from SED has been reported 
\cite{dlPena2012,Franca2012} including ``truly quantum'' properties like the photoelectric effect.
A study of radiative corrections opens up the contact with quantum electrodynamics \cite{Cetto2012}.
A parodox between initial classical and final quantum behaviour in scattering experiments is explained \cite{TheoPullBack}.
Finally, it is realised that the energy throughput of SED to keep matter stable (energy from stochastic fields in, raditation out)
can be seen as an arrow of time, called the subquantum arrow of time, which is more fundamental than the entropic and
cosmological arrows of time \cite{NVienna2014}.

SED is known to work well for harmonic (linear) problems, see e.g. \cite{delaPenaCettobook}.
A fundamental test for nonlinear problems is the ground state of the hydrogen atom. 
While Claverie and Soto put forward that the motion is non-recurrent, and can not have a stationary distribution \cite{ClaverieSoto1982},
Puthoff puts forward a comparison between energy gain and loss terms, concluding that it should be stable \cite{Puthoff1987}.

The theory has been tested numerically on the hydrogen ground state in a 2-$d$ approximation by Cole and Zou 2003, who observed 
an encouraging agreement with the result from the quantum theory \cite{ColeZou2003}. 
New simulations were carried out recently by us (ref. \cite{NLiskaSED1}, to be called NL1) without the 2-$d$ approximation and 
including much more computational detail and power.
However, it was observed that in all runs and all modellings of the system, there occurred self-ionisation.

In quantum theory the contributions from the leading relativistic corrections go as $1/c^2$, hence as $\alpha^2$, 
with $\alpha=1/137$ the fine-structure constant. 
The SED correction to the ground state energy is related to the Lamb shift, which is of order $\alpha^3\log\alpha$.
The relativistic corrections, mechanical terms, are formally more important than the SED effects,
whereas in SED theory they should only act as small contributions to the non-relativistic Hamiltonian.	 
To see whether this paradox plays a role in the final stability question, their role has to be analysed numerically.
For consistency, the stochastic electromagnetic vector potential has to be calculated to relative order $\alpha^2$ as well, 
that is to say, it has to include spatial terms of order $\vr$ and  $\vr^2$.

We take up this challenge and compare the results with the conjecture for the phase space density of the ground state 
 \cite{NRelHatom06}. This theory will be recalled in section 2. Next the electromagnetic fields, Gaussian sums over $3d$ $k$-space,
  are replaced by Gaussian sums over frequency that reproduce the same correlation functions.
   This essential simplification allows us to carry out simulation of the dynamics, which we program in OpenCL.
 The setup is explained in section 4 and the results are reported in section 5. 
In section 6 we close with a discussion.

\section{Relativistic corrections in the hydrogen problem}

The spin-orbit interaction reads 
for an electron of mass $m$ and charge $-e$ in the field of a nucleus with charge $Ze$ ($e>0$),  with Bohr magneton $\mu_B=e\hbar/2m$ in SI  units,

\BEQ 
H_{SO}=\frac{\mu_B}{\hbar m ec^2r}\frac{\d V}{\d r}\vL\ccdot\vS=
\frac{Ze^2}{8\pi\epsilon_0m^2c^2\, r^3}\,\vL\ccdot\vS
\EEQ

The relativistic Hamiltonian including the Darwin term reads

\BEQ
H_{\rm rel}=\sqrt{m^2c^4+p^2c^2}-\frac{Ze^2}{4\pi\epsilon_0r}+\frac{Ze^2}{8\pi\epsilon_0m^2c^2}\frac{\vL\ccdot\vS}{r^3}
+\frac{\hbar^2}{8m^2c^2}4\pi \frac{Ze^2}{4\pi\epsilon_0}\delta(\vr)\EEQ
The Bohr units

\BEQ a_0=\frac{\hbar}{Z\alpha mc}
,\qquad \tau_0=\frac{1}{\om_0}=\frac{\hbar}{Z^2\alpha^2 mc^2},\qquad 
\EEQ
allow to introduce dimensionless variables by $\vr\to a_0\vr$ and $\vp\to (ma_0/\tau_0)\,\vp$,  implying that $\vL=\vr\times\vp\to \hbar\vL$, and
to take consistently $\vS\to\hbar\vS$.
Keeping $|\vS|=\half\sqrt{3}$ one arrives at
\BEQ
\frac{H_{\rm rel}}{mc^2}=\sqrt{1+Z^2\alpha^2p^2}-\frac{Z^2\alpha^2}{r}+\frac{Z^4\alpha^4}{2}\frac{\vL\ccdot\vS}{r^3}+Z^4\alpha^4\frac{\pi}{2}\delta(\vr)
\EEQ
So the dimensionless nonrelativistic  Hamiltonian $H=(H_{\rm rel}-mc^2)/Z^2\alpha^2mc^2$ picks up the leading relativistic corrections

\BEQ
H=\half p^2-\frac{1}{r}-\frac{1}{8}Z^2\alpha^2p^4+\frac{Z^2\alpha^2}{2}\frac{\vL\ccdot\vS}{r^3}+Z^2\alpha^2\frac{\pi}{2}\delta(\vr).
\EEQ
From $\dot\vr=\p_\vp H$ and $\dot\vp=-\p_\vr H$ the dynamics reads for $\vr\neq\vnul$

\BEQ
\label{rdot}
\dot\vr&=&(1-\frac{Z^2\alpha^2}{2} p^2)\vp+\frac{Z^2\alpha^2}{2} \frac{\vS\times\vr}{r^3},\qquad  \\
\dot\vp&=&-\frac{\vr}{r^3}+\frac{Z^2\alpha^2}{2} \frac{\vS\times\vp}{r^3}+\frac{3Z^2\alpha^2}{2} \frac{\vL\cdot\vS}{r^5}\, \vr,
\label{pdot}
\EEQ
while the spin progresses as

\BEQ
\label{Sdot}
\dot\vS
=\frac{Z^2\alpha^2}{2}\frac{\vL\times\vS}{r^3}
=\frac{Z^2\alpha^2}{2} \frac{(\vp\vr-\vr\vp)\ccdot\vS}{r^3} .
\EEQ
It is a standard exercise to demonstrate that the Hamiltonian $H$ and the total angular momentum $\vJ=\vL+\vS$ are conserved, 
here up to terms of order $\alpha^4$.

An external electromagnetic field is added by the minimal substitution $\vp\to\bar\vp= \vp+\beta\vA(Z\alpha\vr,t)$, with $\beta$ given below
and the spatial scale factor $Z\alpha$ expressing the ratio of the Bohr radius to the wavelength of a photon
with Bohr energy $\alpha^2mc^2$. From Eq. (\ref{Sdot}) it is  confirmed that $\vS^2$ remains conserved, as desired.

Stochastic Electrodynamics leads to a specific stochastic electric and magnetic field, as well as to an $\dddot\vr$ damping term, 
see NL1 and Ref.  \cite{delaPenaCettobook}. When we neglect terms of order $\alpha^{7/2}$ higher, 
the contribution $\nabla_i \bar \vp$ to (\ref{pdot})
and the time derivative of (\ref{rdot}) lead to the Abraham-Lorentz or Brafford-Marshall equation for a particle with spin,

\BEQ
\label{rddot=}
\ddot\vr=&-&\frac{\vr}{r^3}-\beta (\vE+\dot\vr\times\vB)+\beta^2\dddot\vr 
+Z^2\alpha^2\frac{\dot\vr^2\vr+2\dot\vr\ccdot\vr\,\,\dot\vr}{2r^3}
\nn\\
&+&\frac{Z^2\alpha^2}{2r^3} (\vS\times\dot\vr- 3\frac{\vr\ccdot\dot\vr}{r^2}\vS\times\vr+3 \frac{\vS\cdot\vL}{r^2}\vr\, )
,\qquad
\EEQ
where  both the fluctuations and the damping involve the small parameter 

\BEQ \label{beta=}
\beta=\sqrt{\frac{2}{3}}Z\alpha^{3/2}=\frac{Z}{1964.71}
, \qquad  \alpha=\frac{e^2}{4\pi\epsilon_0\hbar c}=\frac{1}{137.036}
\EEQ
(In order to have $\beta$ also as prefactor of $\vE$, we absorb a factor $\sqrt{3/2}$ in  $\vA$, $\vB$ and $\vE$.\label{fntd})
We employ the standard rule
${\d\vA}/{\d t}={\p\vA}/{\p t}+\dot\vr\ccdot\nabla\,\vA$ and definitions $\vE=-\p_t\vA$, $\vB=\nabla\times\vA.$
Here $\vE$ is the transversal part of the electric field, next to the longitudinal part, viz., the Coulomb force $-\vr/r^3$.

\section{Representation of the stochastic fields}

In the Coulomb gauge the SED vector potential of a cube of volume $V$ is a sum of plane waves with random coefficients,    

\renewcommand{\vn}{{\bf k}}

\BEA \label{A=ksum}
\vA(\vr,t)= \sum_{\vk,\lambda}
\sqrt{\frac{\hbar}{2\epsilon_0\omega_\vn V}}\, e^{-\om_\vn\tau_c/2}
\hat\eps_{\vn\lambda}&[& A_{\vn\lambda}\sin(\vk \cdot\vr-\om_\vn t) \nn\\
&+&B_{\vn\lambda}\cos(\vk \cdot\vr-\om_\vn t)] 
\EEQ
The wave vector components $k_a=2\pi n_a/V^{1/3}$ involve integer $n_a=\minfty,\cdots,\infty$, ($a=1,2,3$).
The $\hat\eps_{\vn\lambda}$ with $\lambda=1,2$ are polarisation vectors.
The $A_{\vn\lambda}$ and $B_{\vn\lambda}$ are independent random Gaussian variables with average zero and unit variance.
For each ($\vk,\lambda$) term the average energy  $\int_V\d^3r(\frac{\epsilon_0}{2}\langle\vE^2\rangle+\frac{1}{2\mu_0}\langle\vB^2\rangle)
=\half\hbar\om_\vn\,\exp(-\om_\vn\tau_c)$ 
is equal to the photon zero point energy combined with an exponential cutoff; we choose $\tau_c=\hbar/mc^2$, the Compton time of the electron.

\subsubsection{Noise correlators}

Because it is not often listed in text books, 
we start with deriving the autocorrelator of the $\vA$ field $\vC_A(\vr,t;\vq,s)=\langle\A(\vr,t)\A(\vq,s)\rangle$.
It is translationally invariant in space and time, viz. $\vC_A(\vr,t;\vq,s)=\vC_A(\vr-\vq,t-s)$ with

\BEA \vC_A(\vr,t)&=&\frac{\hbar}{2\epsilon_0c}
\sum_\vn \frac{\veen-\hat\vk\hat\vk}{V k} e^{-\om_\vn\tau_c}
\cos(\vk\cdot\vr-\om_\vn t).
\EEA
Symmetry considerations yield the decomposition

\BEQ \label{CA=C0C1}
\vC_A=C_0\veen-C_1\hat\vr\hat\vr,\qquad
{\rm tr} \,\vC_A=3C_0-C_1,\qquad {\rm tr}\,\vC_A\cdot \hat\vr\hat\vr=C_0-C_1.
\EEQ
Let us introduce $C_s$ and $C_p$ as

\BEQ 
C_s=\half\tr \vC,\qquad C_p=\half
\tr\hat\vr\hat\vr\cdot \vC,
\EEQ
which determine

\BEQ\label{C01=}
C_0=C_s-C_p\qquad C_1=C_s-3C_p.\qquad
\EEQ
Integration over angles $\hat \vk$ yields

\BEA 
C_s(\vr,t)&=&
\frac{\hbar}{4\pi^2\epsilon_0c}
\int_{t_0}^\infty\d k\,ke^{-\gam c k} \frac{\sin kr}{ kr} \cos ckt =
-\frac{\hbar}{4\pi^2\epsilon_0c}
\,\Re\frac{1}{\sigma^2-r^2}, \qquad 
\\ \qquad 
C_p(\vr,t)&=&
\frac{\hbar}{4\pi^2\epsilon_0c}
\int_{t_0}^\infty\d k\,ke^{-\gam k}
\frac{\sin kr- kr\cos kr}{( kr)^3}\cos ckt=
\frac{1}{r^2}(F_1-F_2). 
\EEQ
where $\sigma=c(t-i\gam)$.  The functions

\BEQ F_1&=&
\frac{\hbar}{4\pi^2\epsilon_0c}
\int_{t_0}^\infty\d k\,ke^{-\gam c k}
\frac{1-\cos kr}{ k^2}
\cos ckt,\qquad  \nn\\
 F_2&=&
 \frac{\hbar}{4\pi^2\epsilon_0c}
\int_{t_0}^\infty\d k\,ke^{-\gam c k}
\frac{kr-\sin kr}{ k^3r}\cos ckt,
\EEQ
can be found from $C_s$ since $\p_rF_1=\p_r^2(rF_2)=rC_s(r)$, while $F_1\sim F_2\sim r^2$ for $r\to0$. They read

\BEQ F_1&=&
\frac{\hbar}{8\pi^2\epsilon_0c}
\Re\log\frac{\sigma^2-r^2}{\sigma^2},\quad \nn\\
F_2&=&
 \frac{\hbar}{8\pi^2\epsilon_0c}
\Re\left [\frac{\sigma}{r}\log\frac{\sigma+r}{\sigma-r}+\log\frac{\sigma^2-r^2}{\sigma^2}-2\right],
\EEQ
so that there results the spherically symmetric expression

\BEQ
C_p(\vr,t)=
-\frac{\hbar}{4\pi^2\epsilon_0c}\,\frac{1}{r^2}
\,\Re\,\left [\frac{c(t-i\gam)}{2r}\log\frac{c(t-i\gam)+r}{c(t-i\gam)-r}-1\,\right].
\qquad
\EEA

\subsection{Small distance corrections to the fields and correlators}

\newcommand{\bq}{\bar q}
\newcommand{\br}{\bar r}
\newcommand{\bvq}{\bar{\bf q}}
\newcommand{\bvr}{\bar{\bf r}}
\newcommand{\vomn}{}
\newcommand{\np}{m}
\newcommand{\Ain}{A_{ni}}
\newcommand{\Bin}{B_{ni}}

Our aim is to determine a numerically treatable presentation of the stochastic fields beyond the dipole (i.e., $r\to0$) approximation of NL1.
We switch from SI units to Bohr units. This implies that $e\vA$ is replaced by $\beta\mathring{\vA}$, with $\beta$ defined in (\ref{beta=}) and $\mathring{\vA}$
 being the sum (\ref{A=ksum}) with the argument of the square root replaced by $3\pi/\mathring{V}\mathring{k}$, 
 involving the normalised variables $\mathring{V}=V/(c\tau_0)^3$ and  $\mathring{k}=\om \tau_0$.
 In these units it holds that $\tau_c=Z^2\alpha^2\ll1$, while $C_s$ and $C_p$ read

 \BEA 
C_s(r,t)&=& -\frac{3}{2\pi}\,\Re\frac{1}{(t-i\gam)^2-\br^2}, \qquad \qquad  
\nn\\
C_p(r,t)&=&-\frac{3}{2\pi \br^2}\,\Re\,\left [\frac{t-i\gam}{2\br}\log\frac{t-i\gam+\br}{t-i\gam-\br}-1\,\right],
\qquad
\EEQ
to be taken at $\br=Z\alpha r$.  Clearly, we need only small $\br$, where these functions have leading order behaviours 
 \BEA 
C_s(r,t)&=& -\frac{3}{2\pi}\,\Re\left[\frac{1}{(t-i\gam)^2}+\frac{\br^2}{(t-i\gam)^4}\right], \qquad \nn\\
C_p(r,t)&=&-\frac{1}{2\pi}\,\Re\,\left [\frac{1}{(t-i\gam)^2}+\frac{3\br^2}{5(t-i\gam)^4}\right].
\EEQ
From (\ref{CA=C0C1}) and (\ref{C01=}) we collect the leading behaviour for the $\vA$ autocorrelation

\BEQ\label{CAtrs0u0}
\vC_A(\vr,t;\vnul,0)
=-\frac{1}{\pi}\Re\frac{\veen}{( t-i\tau_c)^2}+\frac{3}{5\pi}\Re\frac{\bvr\,\bvr-2\br^2\veen}{( t-i\tau_c)^4}.
\EEQ

Numerically, it is far too cumbersome to deal with the $3-d$ sum over $\vk$ values in the definition (\ref{A=ksum}).
In NL1 it was realised for the limit $\vr\to\vnul$ that, since the $\vA$ fields are sums of Gaussian random variables, 
they themselves are Gaussian and may just as well be replaced by different Gaussian sums that produce the same autocorrelator.
We wish to extend this approach to include the $\vr$-dependence  to quadratic order. Hence we write

\BEQ
C^A_{ij}(\vr,t;\vq,s)=\int_{t_0}^\infty\d\om \,W^2(\om)C^A_{ij}(\vr,\vq,\om) 
\,e^{i\om(t-s)},\qquad\EEQ
with $ W(\om)=e^{-\om\tau_c/2}$, which demands

\BEQ
\hspace{-5mm}
C^A_{ij}(\vr,\vq,\om)=\langle A_i(\om,\vr)A_j(-\om,\vq)\rangle
&=&\frac{\om}{\pi}\delta_{ij}+\frac{\om^3}{10\pi}[\br_i\br_j+\bq_i\bq_j-2(\br^2+\bq^2)\delta_{ij}] 
\nn\\&&
+\frac{\om^3}{10\pi}(4\br_k\bq_k\,\delta_{ij}-\br_i\bq_j-\br_j\bq_i)
\EEQ

Let us recall the solution for the $\vr=\vq=\vnul$ case of NL1. First, the frequency integral is discretised.
For uniformly spaced frequencies $\om_n=n\d\om$ with $n=1,2,3,\cdots$, the leading term may arrive from

\BEQ
\vA(\vnul, t)=\sum_{n=1}^\infty \sqrt{\frac{\om_n\d\om}{\pi}}  W(\om_n)\left(\, \vA_n\sin\om_nt+\vB_n\cos\om_nt\,\right) ,
\EEQ
having independent, real valued Gaussian random variables $\Ain  $ and $\Bin  $ for $n=1,2,3,\cdots$ and $i=1,2,3$, with zero average and unit variance.


The new task is to extend this for small $\vr$ and $\vq$.
 Since the autocorrelator has a correction of order $r^2$, we want for $\vA$ linear and bilinear terms.
While the leading correction must be linear in $\vr$ or $\vq$, it happens that 
the term of $\vA$ bilinear in $\vr$ can be determined explicitly. 
This leads to the component of $\vA$

\BEQ
A_i(\om_n,\vr)=\Ain +b_{nik}\br_k
+\frac{\om^2}{10}[\br_ia_k\br_k-2a_i\br^2],
\qquad
\EEQ
for $i=1,2,3$ and with sum over $k=1,2,3$ implied.

To consider the terms linear in $\vr$, we introduce 8 real valued matrices, which coincide up to a factor with the generators of SU(3),
\BEQ
\hspace{-1cm}
\bar\lambda^1&=&\sqrt{3}\left(\begin{tabular} {@{}lll}  0 & 1 & 0 \\ 1 & 0 & 0 \\ 0 & 0 & 0 \end{tabular} \vspace{-2mm} \right),\qquad
\bar\lambda^2=\sqrt{5}\left(\begin{tabular} {@{}lll}  0 & -1 & 0 \\ 1 & 0 & 0 \\ 0 & 0 & 0 \end{tabular}\vspace{-2mm}\right),\qquad 
\nn \\ \hspace{-1cm}
\bar\lambda^3&=&\sqrt{3}\left(\begin{tabular} {@{}lll}  1 & 0 & 0 \\ 0 & -1 & 0 \\ 0 & 0 & 0 \end{tabular}\vspace{-2mm}\right),\qquad
\bar\lambda^4=\sqrt{3}\left(\begin{tabular} {@{}lll}  0 & 0 & 1 \\ 0 & 0 & 0 \\ 1 & 0 & 0 \end{tabular}\vspace{-2mm}\right),\qquad   \nn\\
\hspace{-1cm}
\bar\lambda^5&=&\sqrt{5}\left(\begin{tabular} {@{}lll}  0 & 0 & -1 \\ 0 & 0 & 0 \\ 1 & 0 & 0 \end{tabular}\vspace{-2mm}\right),\qquad
\bar\lambda^6=\sqrt{3}\left(\begin{tabular} {@{}lll}  0 & 0 & 0 \\ 0 & 0 & 1 \\ 0 & 1 & 0 \end{tabular}\vspace{-2mm}\right),\qquad 
 \nn \\ \hspace{-1cm}
\bar\lambda^7&=&\sqrt{5}\left(\begin{tabular} {@{}lll}  0 & 0 & 0 \\ 0 & 0 & -1 \\ 0 & 1 & 0 \end{tabular}\vspace{-2mm}\right),\qquad
\bar\lambda^8=\hspace{5mm}\left(\begin{tabular} {@{}lll} 1 & 0 & 0 \\ 0 & 1 & 0 \\ 0 & 0 & -2 \end{tabular}\vspace{-2mm}\right).
\EEQ 
At each frequency $\om_n>0$ 
we express the random $b_{nik}$ in terms of a scale $C_n$ and 8 independent, complex valued Gaussian random variables $\beta_{na}$,

\BEQ
b_{nik}\vomn =\sqrt{C_n\vomn }\sum_{a=1}^8 \beta_{na}\vomn \bar\lambda^a_{ik}.\qquad 
\EEQ
The $\beta$'s are chosen to have averages zero and unit variances in the manner

\BEQ 
&& \beta_{na}\vomn =\frac{1}{2}[\beta_{na}^{(1)}\vomn -i\beta_{na}^{(2)}\vomn],\qquad 
\langle \beta_{na}^{(1)}\vomn \beta_{\np b}^{(1)}\rangle=\langle \beta_{na}^{(2)}\vomn \beta_{\np b}^{(2)}\rangle=\delta_{ab}\delta_{n\np }, \qquad
\EEQ
while by assumption $\langle \beta_{na}^{(1)}\rangle=\langle \beta_{na}^{(2)}\rangle=\langle \beta_{na}^{(1)}\beta_{\np b}^{(2)}\rangle=0$.
This implies that

\BEQ
\langle \beta_{na}\vomn \beta_{\np b}^\ast\vomn \rangle=\half\delta_{ab} \delta_{n\np }\qquad (a,b=1,\cdots, 8),
\EEQ
while a bit of algebra shows that

\BEQ
\sum_{a=1}^8
\bar\lambda_{ik}^a\bar\lambda^a_{jl}=2(4 \delta_{ij}\delta_{kl} - \delta_{ik} \delta_{jl} - \delta_{il} \delta_{jk}\,).
\EEQ
Hence the choice 
\BEQ
C_n=\frac{\d\om\,\om_n^3}{20\pi}
\EEQ
reproduces at each $n$ the desired cross terms of $\vr$ and $\vq$,

\BEQ
\langle b_{nik}b_{njl}^\ast\rangle \br_k\bq_l=\frac{\d\om\,\om_n^3}{20\pi} \,(4\br_k\bq_k \delta_{ij} -\br_i\bq_j-\br_j\bq_i).
\EEQ
with sums over $k$ and $l$ implied. 
We end up with the field up to order $r^2$,

\myskip{\BEQ\label{vAr=}
 A_i(t,\vr)&=&\sum_{n>0}\sqrt{\frac{\d\om\,\om_n}{\pi}}\{ \Bin  \cos\om_nt +\Ain  \sin\om_nt \}
+\sum_{n>0}\sqrt{\frac{\d\om\,\om_n^3}{20\pi}}
\{\beta^{(1)}_{na}\cos\om_nt +\beta_{na}^{(2)}\sin\om_nt \}\bar\lambda^a_{ik}\br_k  \nn \\ 
&+& 
\sum_{n>0}\sqrt{\frac{\d\om\,\om_n^5}{100\pi}}\, \{[ B_{nk}\br_k\br_i-2\Bin   \br^2]\cos\om_nt +
[A_{nk}\br_k\br_i-2\Ain  \br^2]\sin\om_nt \} 
\EEQ}

\BEQ\label{vAr=}
 A_i(t,\vr)&=&\sum_{n>0}\sqrt{\frac{\d\om\,\om_n}{\pi}}W(\om_n)\left\{
 ( \Bin  \cos\om_nt +\Ain  \sin\om_nt ) 
\right. \nn\\ &+& 
 \frac{\om_n}{4\sqrt{5}} (\beta^{(1)}_{na}\cos\om_nt +\beta_{na}^{(2)}\sin\om_nt )\bar\lambda^a_{ik}\br_k 
 \\  &+& \frac{\om_n^2}{10}
 [( B_{nk}\br_k\br_i-2\Bin   \br^2)\cos\om_nt +
(A_{nk}\br_k\br_i-2\Ain  \br^2)\sin\om_nt ]\} \nn
\EEQ
summed over $k=1,2,3$ and $a=1,\cdots,8$. Notice that the Coulomb gauge condition $\p_iA_i=0$ is satisfied and that
the translation invariant autocorrelator (\ref{CAtrs0u0}) comes out right up to order $r^2$ included.

\myskip{
\BEQ
\langle A_i(t,\vr)A_j(s,\vq)\rangle
&=&\sum_n\frac{\d\om\,\om_n}{\pi}\cos\om_n(t-s)\left\{\delta_{ij}
+\frac{\om_n^2}{20}\sum_{a=1}^8\bar\lambda_{ik}^a\bar\lambda^a_{jl}\br_k\bq_l+\frac{\om_n^2}{10}[\br_i\br_j+\bq_i\bq_j-2\delta_{ij}(\br^2+\bq^2)] \right\} \nn \\
&=&\sum_n\frac{\d\om\,\om_n}{\pi}W(\om_n)\cos\om_n(t-s)\left\{\delta_{ij}
+\frac{\om_n^2}{10}[(\br_i-\bq_i)(\br_j-\bq_j)-2\delta_{ij}(\bvr-\bvq)^2 ]\right\} ,
\EEQ
which is translationally invariant. We have reinserted the cutoff function $W$. For $W(\om)=e^{-\om\tau_c/2}$ it leads to
(\ref{CAtrs0u0}).}

We now acknowledge that the right hand side of  (\ref{vAr=}) should be taken at $\bvr=Z\alpha \vr$.
The electric field $E_i(\vr,t)=-\p_tA_i(\vr,t)$ thus follows as

\BEQ\label{Eir=}
E_i(\vr,t)&=&
\sum_{n>0}\sqrt{\frac{\d\om_n\om_n^3}{\pi}} W(\om_n)\{ (B_{ni}\sin\om_nt -A_{ni}\cos\om_nt ) \nn\\
&+&Z\alpha \frac{\om_n}{4\sqrt{5}}
 (\beta^{(1)}_{na}\sin\om_nt -\beta_{na}^{(2)}\cos\om_nt )\bar\lambda^a_{ik}r_k 
\\&+&  Z^2\alpha^2\frac{\om_n^2}{10} [ \,(B_{nk}r_kr_i-2B_{ni} r^2)\sin\om_nt -(A_{nk}r_kr_i-2A_{ni}r^2)\cos\om_nt \,] \}. \nn
\EEQ
As is well known \cite{delaPenaCettobook,ColeZou2003}, the $\vB$ field is of order $Z\alpha$. Indeed, the field tensor
$ F_{ij}=\p_iA_j-\p_jA_i$ reads

\BEQ\label{Fijr=}
&& F_{ij}= -Z\alpha \sum_{n>0}
\sqrt{\frac{\d\om_n\om_n^3}{5\pi}}\,W(\om_n)\sum_{a=2,5,7} (\beta^{(1)}_{na}\cos\om_nt +\beta_{na}^{(2)}\sin\om_nt \,)\bar\lambda^a_{ij } +   \\
&&
Z^2\alpha^2 \sum_{n>0}\sqrt{\frac{\d\om_n\om_n^5}{4\pi}}\,W(\om_n) [\,(B_{ni}r_j-B_{nj}r_i)\cos\om_nt + (A_{ni}r_j- A_{nj}r_i)\sin\om_nt ],\nn
\EEQ
containing only the antisymmetric $\bar\lambda^a$.
The magnetic field components follow as $B_k=\half \eps_{ijk}F_{ij}$ 
and the Lorentz force reads $(\dot\vr\times\vB)_i = 
F_{ij}\dot r_j$.

With the electric and magnetic fields expressed as single sums over the frequency, we can now implement this structure in the numerical code. 
In our OpenCL method  not much extra time is demanded for handling at each $\om_n$ the 16 additional random variables $\beta^{(1,2)}_{na}$.

\section{Numerics}

In our code we numerically solved the ODE in (9) for a charge of $Z=3$  using a fourth order Runge-Kutta scheme.  
For our most important run, we cut off the frequency in equations (36) and (37) using a `moving' cut-off at 2.5 times the Keplerian frequency of the electron. 
The used code is an enhanced version of the code described in [12]. 
This code was/is highly optimized using OpenCL and it is run on a state-of-the art GPU (AMD R9 290X). 
The present performance is equivalent to a CPU-only version of the code running on a 300 core CPU cluster.
Here we give a pointwise summary of the main features/caveats and the differences to the version described in ref.  [12]:

1) 
The considered frequencies are $\om_n=n/N$. 
Due to the presence of a vast amount of random variables we had to set  $N=10^5$ instead of  $N=10^6$ as was done in our previous article [12].  
Up to $t=N$ the autocorrelator of equation (35) reduces to the autocorrelator (23) of the exact $\vA$ field, defined in equation (11). 
For larger $t$, we may not simulate a genuine $3D$ problem anymore.
This could be a problem since it is shorter than the damping time $1/\beta^2$ and possibly also shorter than other timescales relavant to
 the electron's dynamics.

2) We used the 'classical' Runge Kutta fourth order integration scheme with full energy 
and angular momentum conservation up to eccentricities of 0.99 and higher. We did also test other ODE schemes
like several of the Adam-Bashforth type and the simpler Euler method. On top of that we did vary the number of iterations per orbit from run to run. 
Thus we can conclude that our results are ODE scheme independent, with the biggest numerical error induced due to the other points mentioned in this paragraph.

3) We used 4000 iterations per orbit. 
The EM-field, though, was only updated 25-30 times per period of the highest mode with the other points determined by higher order interpolation using Lagrange Polynomials. 
We calculated this  EM-field 1.5 to 2 times as often with respect to our previous article [12], since higher numerical precision was required so to assure that the effects 
of the relativistic corrections, the magnetic field, the positional dependence of the EM-field and the spin-orbit coupling were fully accounted for.

4) As was done in our previous article, we updated the moving cut-off frequency when the orbital period of the electron changed more than 20\%. 
This was done to minimise the presence of discontinuities in the equation of motion (9). 
Note that this field switching could still  induce errors in our solution.

5) The code spends a very long time to simulate the electron at low energies. 
To speed the code up, we gave the electron a push to increase its energy when it became smaller than $\cE=-1.6$.
This is justified by the resulting energy distribution and the earlier published conjecture [13].
For the present article we chose a scheme in which we randomly gave the electron a push either parallel or perpendicular to its velocity axis. 
The advantage of this scheme is that it remains simple and induces less bias in the angular momentum at low energies.

6) We made use of the Runge Kutta RK4 algorithm. This is not the main source for our errors,  because others issues like the switching of the EM field every few hundred steps, 
the finite cutoff frequency and other points mentioned above, produce (much) bigger errors. We did do extensive convergence testing, though, even runs with full 
double precision and 4 times as many steps as used in the reported results (but with smaller $N$ values). No statistical difference was detected in any 
alternative simulation. 
On top of that,  we experimented with several Adam Bashforth like methods, again no difference was detected at all. We could even have used the standard
 Euler method. Overall, we have more than 30 runs in different configurations and with slightly varying parameters, all giving approximately the same results.
Since we only investigate the statistical behaviour of SED, but do not claim to solve it with high precision at a very late times, 
we are convinced of the fact that the reported results give a fair presentation of the situation.

\section{Results}
We ran the code several times using different values for the cut-off frequency. The longer runs took 50 to 200 GPU hours to complete.

Here we present the results of our most promising run up to the point that the electron energy went above $\cE$=-0.05, what we define as  ionisation. 
The moving cutoff was set at 2.5 times the electron's Keplerian frequency. 
In figures 1a,1b we show the time series for the energy and radius while in figures 2a, 2b we show the corresponding probability distributions 
compared to the 1s ground state and an earlier published conjecture [13].
This conjecture for the classical distribution of the energy  $\cE$ and the angular momentum $L$ reproduces the  quantum 1s ground state results.
In table 1 we give the run time of this simulation before ionisation and we compare it to the damping timescale of 1/$\beta^2$.

\begin{center}     
\noindent     
\begin{tabular}{||r|r|r||}
 \hline \hline    
property & value & duration (s) \\ 
 \hline\hline
$ t_{total}$ & $2.05\,10^7\,t_0$ & $5.5\,10^{-11}$ s \\ \hline
$t_{damp}$&  $4.28\,10^5\,t_0$ & $1.15\,10^{-12}$ s \\ \hline
$N_{orbit}$ &  $3.26\,10^6$ &  \\ \hline
$N_{damp}$ & 48 & \\ \hline \hline
\end{tabular}     
\end{center}     
\centerline{Table 1: Time and number of orbits for our simulation with  2.5 harmonics and $Z=3$.}

\begin{figure}[h]
\includegraphics[width=5cm]{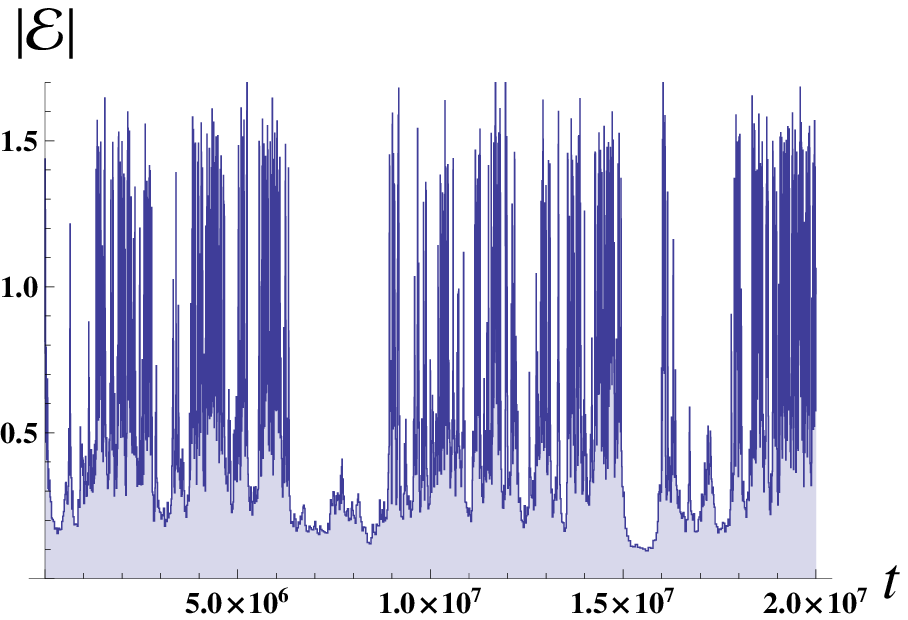} \hspace{1cm}  \includegraphics[width=5cm]{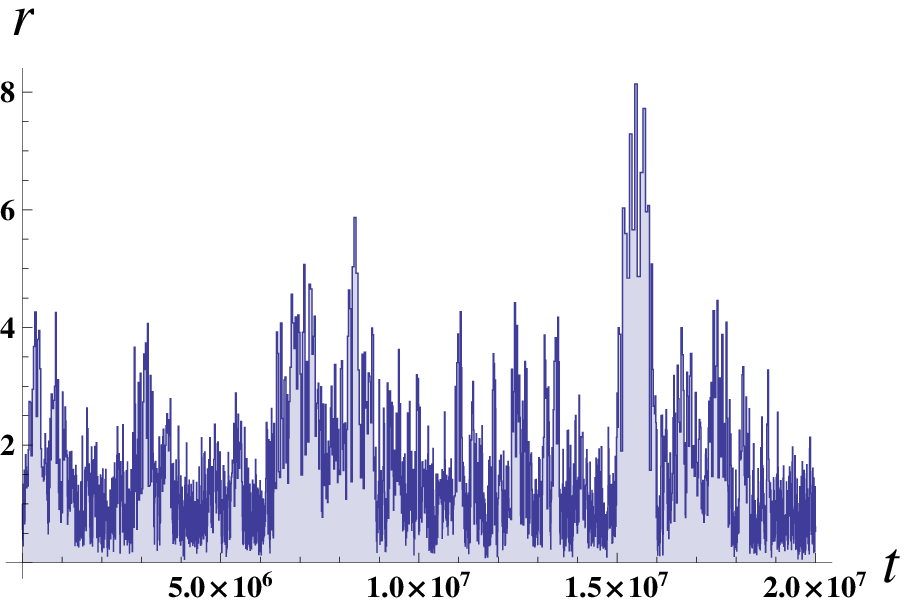}
\caption{a): Time series for the energy, in Bohr units. b): Time series for the radius.} 
\label{fig:1}    
\end{figure}


\begin{figure}[h]
\includegraphics[width=5cm]{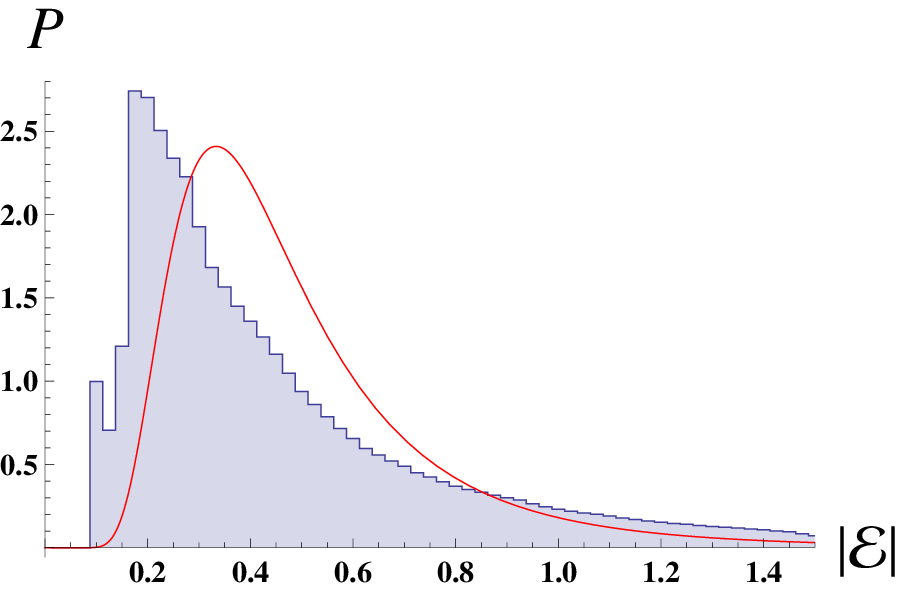} \hspace{1cm}   \includegraphics[width=5cm]{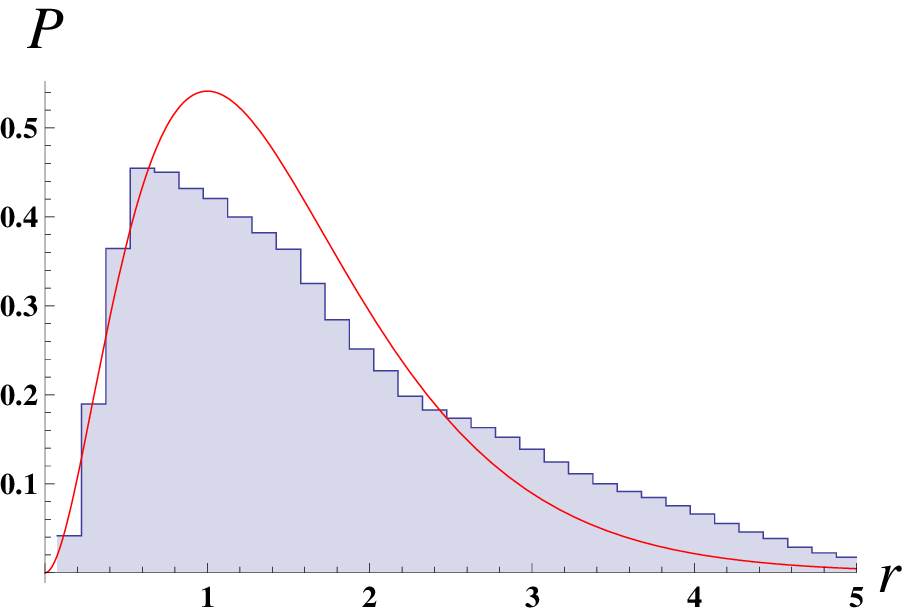}
\caption{a): Histogram for the energy data  of Fig. 1a. The curve is the conjecture
$P(\cE )=(4/3|\cE |^6) \exp(-2/ |\cE |)$ from refs. \cite{NRelHatom06,NLiskaSED1}.
b): Histogram for the radius data  of Fig. 1b. The curve is the quantum prediction $P=4r^2\exp(-2r)$.} 
\label{fig:3}      
\end{figure}


These results suggested that the electron remained stable and had a probability distribution closer to the conjecture [12, 13] for a longer time than was presented in our previous article [12]. 
However, we did find out that the time up until the electron's ionisation varied quite drastically between $t=10^6 t_{0}$ to $t=10^7 t_{0}$ for different runs using either the old code or the new code. 
Thus we can conclude that there is no statistically significant difference between this run and the idealised run in our previous article [12].
Runs with the moving cutoff set at 4.5 or 6.5 times the orbital frequency showed ionisation on even shorter timescales. 
Multiple runs with a fixed cutoff set at 1.5 times 
the Kepler frequency for $\cE=-1.6$ did even show ionisation within $t=10^4 t_{0}$, though the accuracy of the interpolation scheme may be 
brought into question for the fixed cut-off case, especially when we want to account for the higher order effects discussed in this article. 
This is because for high electron energies very strong EM-modes are included, the frequency of which is $30$ 
times the Keplerian frequency of the electron, while we expect only the first few `harmonics' to have a significant contribution [11, 12]. 
Therefore the numerical error in the value of the EM-field can become as large as 20 percent of the first harmonic.

\section{Discussion}
We have extended our previous dynamical simulation for the hydrogen ground state in Stochastic
Electrodynamics (SED) to include relativistic corrections: the $p^4$ kinetic correction and the spin-orbit coupling.
Formally these terms are larger than the Lamb shift, which should be responsible for effective quantum
behaviour of SED.  While in our previous approach \cite{NLiskaSED1} a self ionisation was observed,
the question is how the relativistic corrections -- stronger than the SED corrections -- impact on this.
Hereto the stochastic fields, sums over $3d$ $k$-space, are first expressed as frequency sums of
stochastic variables that reproduce the same correlations up to the needed order $\vr^2$. 
Next, the routine is implemented in OpenCL.

The answer to the stability question is negative: the relativistic corrections have hardly any impact: self ionisation remains present;
the relativistic corrections just act as small terms to the non-relativistic Hamiltonian.
This is at odds with a prediction put forward by Boyer  \cite{Boyer2003,Boyer2004},
who argues in favor of a relativistic treatment to arrive at the quantum result for the H atom. 
This is now shown not to work out.
The continued finding of self-ionisation underlines the correctness of leaving out the relativistic corrections at
the most basic level of the theory, as is indeed routinely done in the field. 
The observed discrepancy between SED and quantum mechanics may also may carry ingredients of the 
SED description of the free particle \cite{NLiskaSED1,PenaFreeParticle}.

Whether the thus observed self ionisation is the final nail in the coffin of SED, or, perhaps, due to 
approximations made, remains an open question. The fact that ionisation does not happen (in the employed window approximation) 
at early times but at some later time
is likely due to the fact that the electron has to find the niche in which it can ionise and stay there sufficiently long.
An issue of special interest is the point charge approximation for the electron.
We stress in this respect that the point particle theory is incapable
to deal with the Darwin term of Eq. (2), a relativistic correction proportional to a delta-function.
In this sense, the employed theory could not have been correct from the start.
Whether these deficiencies can be simultaneously overcome remains a challenge for the future.

\begin{acknowledgements}
It is a pleasure to thank Erik van Heusden for discussion.
\end{acknowledgements}

\section*{Conflict of interest}
The authors declare that they have no conflict of interest.



\end{document}